\documentclass[aip,apl,amsmath,amssymb,reprint,superscriptaddress]{revtex4-1}
\usepackage{graphicx}
\usepackage{amsmath}
\usepackage{dcolumn}
\usepackage{bm}
\usepackage{bbold}
\usepackage{relsize}

\bibstyle{apsrev4-1}

\begin{document}
    
\title{Aggregate Frequency Width, Nuclear Hyperfine Coupling and Jahn-Teller Effect of $Cu^{2+}$ Impurity Ion ESR in $SrLaAlO_4$ Dielectric Resonator at $20$ Millikelvin}

\author{M. A. Hosain}
\email{akhter361@yahoo.co.uk}
\affiliation{ARC Centre of Excellence for Engineered Quantum Systems, School of Physics, University of Western Australia, 35 Stirling Highway, Crawley WA 6009, Australia.}

\author{J.-M. Le Floch}
\affiliation{MOE Key Laboratory of Fundamental Physical Quantities Measurement, School of Physics, Huazhong University of Science and Technology, Wuhan 430074, Hubei, China.}
\affiliation{ARC Centre of Excellence for Engineered Quantum Systems, School of Physics, University of Western Australia, 35 Stirling Highway, Crawley WA 6009, Australia.}

\author{J. Krupka}
\affiliation{Department of Electronics and Information Technology, Institute of Microelectronics and Optoelectronics, Warsaw University of Technology, Koszykowa 75, 00-662 Warszawa, Poland.}

\author{M. E. Tobar}
\affiliation{ARC Centre of Excellence for Engineered Quantum Systems, School of Physics, University of Western Australia, 35 Stirling Highway, Crawley WA 6009, Australia.}



\begin{abstract}

The impurity paramagnetic ion, $Cu^{2+}$ substitutes $Al$ in the $SrLaAlO_4$ single crystal lattice, this results in a $CuO_6$ elongated octahedron, the resulting measured g-factors shows four-fold axes variation condition. The aggregate frequency width of the electron spin resonance with the required minimum level of impurity concentration has been evaluated in single crystal $SrLaAlO_4$ at $20$ millikelvin. Measured parallel hyperfine constants, $A_{\scriptscriptstyle\parallel Cu}$, were determined to be $-155.7\times10^{-4}~cm^{-1},~ -163.0\times10^{-4}~cm^{-1},~ -178.3\times10^{-4}~cm^{-1} $ and$~-211.1\times10^{-4}~cm^{-1}$ at $9.072~GHz~(WGH_{4,1,1})$ for the nuclear magnetic quantum number $M_I=+\frac{3}{2},+\frac{1}{2},-\frac{1}{2}$,~and$-\frac{3}{2}$ respectively. The anisotropy of the hyperfine structure reveals a characteristics of static Jahn-Teller effect. The second-order-anisotropy-term, $\sim (\frac{spin-orbit~coupling}{10D_q})^2$, is significant and can not be disregarded, with the local strain dominating over the observed Zeeman-anisotropy-energy difference. The Bohr electron magneton, $\beta=9.23\times 10^{-24} JT^{-1}$, (within $-0.43\%$ so-called experimental error) has been found using the measured spin-Hamiltonian parameters. Measured nuclear dipolar hyperfine structure parameter $P_{\scriptscriptstyle\parallel}=12.3\times10^{-4}~cm^{-1}$  shows that the mean inverse third power of the electron distance from the nucleus is $\langle r^{-3}_q\rangle\simeq 5.23$ a.u. for $Cu^{2+}$ ion in the substituted $Al^{3+}$ ion site assuming nuclear electric quadruple moment $Q=-0.211$ barn.
      
\end{abstract}

\maketitle

\subsection{Introduction:}

Recently there has been renewed interests and experimental studies devoted to spins in solids due to the emergence of new quantum hybrid systems, which requires the manipulation of spin quantum states$\cite{KurizkiHybSy,RoadMapHybSy,ContHybSy}$, with continued searches for viable candidates$\citep{RoadMapHybSy,Buluta}$. In this work, we implement the whispering gallery $(WG)$ mode technique to study impurity paramagnetic ions unpaired electron spin resonance possessing nuclear hyperfine coupling in a dielectric crystal lattice$\citep{Farr,Karim,Tobar1997}$. Site symmetry information of impurity paramagnetic ions in the $SrLaAlO_4~(SLA)$ single crystal lattice is acquired through $WG$ multi mode $ESR$ spectroscopy ($Fig.\ref{CuDens},\ref{CuAniso1},\ref{CuScratchPlot1}$ and $\ref{CuScratchPlot2}$), providing measurement of hyperfine structure broadening, g-factor variation and other anisotropy effects. 

WG mode spectroscopy is highly sensitive, combined with the multi-mode nature of the experimental results provides values of some fundamental physics quantities with high precision. The Jahn-Teller effect in a metal-ligand octahedral complex, $ML_6$, usually induces charge coupling, orbital and magnetic ordering, displacement, and underlines the structural details in determining electronic behaviour$\cite{JT-Static,JT-HyperBroad,JT-phenomena,JT-visu}$. Such kind of high precision measurements of the hyperfine structure characteristics are essential for quantum state mapping. The unpaired electron spin moment reveals information about the spin-Hamiltonian parameters, including hyperfine structure anisotropy in a single spectroscopic term of anomalous Zeeman transitions when each $WG$ mode of resonance frequency, $\omega$, is equal to the Larmor precession, $\omega_L$, of the magnetic dipoles$\cite{Farr,Karim,Maxim}$. Also, it is possible to determine the difference in the spin-Hamiltonian parameters due to tetragonal distortions using ESR spectroscopy$\cite{sH-CuGaSe,TetraCu-ZCO}$. Luminance and phosphorescence were deteermined in other perovskite crystals with $Cu^{2+}$ defect centres as well$\cite{PreLumSAlO,ProsforSAlO}$. Feng and Zheng$\cite{EPRgCuGaSe}$ gave a theoretical analysis of spin-Hamiltonian parameters for the rhombic $Cu^{2+}$ centres in $CuGaSe_2$ crystal. A suitable low-loss crystal is required for possible use as a host of paramagnetic ions for this $ESR$ study. Sequential excitation of WG modes of X-band to Ku-band frequencies in the low-loss $SrLaAlO_4$ dielectric bulk crystal under systematic control of high resolution vector network analyzer (VNA) with a sophisticated coupled co-axial loop allows us to perform these measurements and analysis. In this dielectric resonance process, the WG modes energy loss mechanisms was minimized achieving a high loaded Q-factor $Q_L$ for this ESR spectroscopy at $20$ millikelvin $(mK)\citep{Krupka,LeFloch,WG,Krupka1}$.\

 High confinement of electromagnetic field in the low-loss crystal $SrLaAlO_4~(SLA)$ ensures high sensitivity of ESR spectroscopy for very low concentration of impurity ions. Implementing this technique, the microwave-power and other terms are kept constant, the required minimum number of impurity ion follows the proportionality\cite{Hartnett1999}$\citep{CharlesESR}$ $N_{min} \propto \frac{1}{\omega Q_L}$ for detection of ESR transition spectrum, and is estimated for $Cu^{2+}$ ion of unpaired electron effective spin $S=\frac{1}{2}$ as:
 \begin{eqnarray}
\label{eq:Nmin}
 N_{min}=\Big(\frac{4k_B V_sT_s}{g_e^2\beta^2\mu_\circ}\Big)\Big(\frac{\Delta \omega}{\omega}\Big)\Big(\frac{1}{\eta Q_L}\Big)\Big(\frac{P_n}{P}\Big)^\frac{1}{2}   
\end{eqnarray} 
 Where $k_B$ is the Boltzman constant, $V_s$ is the mode volume, $T_s$ is the sample temperature, $g_e$ is the electron g-factor, $\beta$ is the Bohr electron magneton, $\mu_\circ$ is the magnetic permeability of free space, $\omega$ is the resonance frequency (or transition energy frequency), $\eta$ is the filling factor, $P_n$ is the noise intensity, $P$ is the microwave input intensity, and $\Delta \omega$ is the width of aggregated spin frequency at resonance which is depended on the 'shape function' $f(\omega)$ normalized as $\int_0^\infty f(\omega)\partial\omega=1$ for a wide range of Larmor precession. Significant output (transmission) occurs only at the resonance in a very narrow frequency width $\Delta \omega$ in the region $\omega\approx\omega_L$ for ESR $(Fig.~\ref{CuDens})$. Previously, many studies devoted to high sensitive measurements of various types of resonators over a wide range of frequencies with a variety of probing system have been performed$\citep{AnninoQL,LongoQL,ColligianiQL,AndersMK,YapMK}$. Benmessai et al.$\citep{Karim}$ described a typical concentration level measurement of impurity $Fe^{3+}$ ion in sapphire using WG mode microwave resonance. Anders et al.$\citep{AndersMK}$ described a single-chip electron spin resonance detector operating at $27~GHz$.\\

    The single crystal $SrLaAlO_4$ crystallizes in a $K_2NiF_4$-structure of space group $I4/mmm$  with the lattice constants of $a=b=3.756$~\r{A} and $c=12.636$~\r{A}, grown by the Czochralski technique$\citep{Anna,GrowthSLA,Woensdregt,Co-inSla}$. The $Al^{3+}$ site is surrounded by a tetragonally elongated (along c-axis) oxygen octahedron of two bonds $Al-O2$ of length $R_\parallel = 2.121$~\r{A} and four coplanar bonds $Al-O1$ of length $R_\perp = 1.885$~\r{A} between aluminium and oxygen$\citep{Co-inSla,DefectCu^2}$. In the substituted $Al^{3+}$ ion site, the $d^9$ ground state of $Cu^{2+}$ ion $\mid x^2-y^2\rangle = \mid2^s\rangle$ orbital singlet of spin doublet (superscript 's' indicates symmetric wave function) is valid to consider as a cubic symmetry with small deformation in the $CuO_6$ octahedral structure of complex due to elongation along z-axis (c-axis) under tetragonal distortion$\citep{PilbrowESR,AbragamESR}$. Hence, an applicable hyperfine spin-Hamiltonian in the z-axis symmetry of ion of nuclear spin $I$ with electron effective spin $S=\frac{1}{2}$ takes the form$\citep{AbragamESR}$:
\begin{eqnarray}
\label{eq:CuSH1}
{\lefteqn { {\mathcal{H}_n}= A_{\scriptscriptstyle\parallel Cu} S_zI_z + A_{\scriptscriptstyle\perp Cu} ( S_xI_x + S_yI_y) +\nonumber}}\\ & & { P_{\scriptscriptstyle\parallel}\bigg\{ I_z^2 - \frac{1}{3} I(I+1) \bigg\} - \beta B {g^I_{\scriptscriptstyle\parallel Cu}}I_z}   
\end{eqnarray} 
 $A_{\scriptscriptstyle\parallel Cu}$ and $A_{\scriptscriptstyle\perp Cu}$ are parallel and perpendicular component of hyperfine constant $\textbf{A}$, the magnetic field is $\textbf{B}$ applied along z-axis and $g^I_{\scriptscriptstyle\parallel Cu}$ is the nuclear parallel g-factor aligned with electronic parallel g-factor ${g_{\scriptscriptstyle\parallel Cu}}$. The dipolar hyperfine structure parameter is considered due to interaction with electric quadrupole moment of nucleus $Q$, which is in this symmetry $P_{\scriptscriptstyle\parallel} = -{\frac{3e^2Q}{7I(2I-1)}}\langle r_q^{-3}\rangle$. Where $\langle r_q^{-3}\rangle$ is the mean inverse third power of the electron distance from the nucleus (origin), averaged over the electronic wave-functions. For the unfilled d-shell paramagnetic unpaired electron, this term is $\langle r^{-3} \rangle = \frac{\langle r_q^{-3} \rangle}{1-R_q}$, where $R_q$ is $0.1$ to $0.2$ approximately.\\
 
 The hyperfine structure lines are separated by an amount of magnetic field $\frac{ A}{g_e \beta}$ (considering $1^{st}$ order perturbation)$\citep{PilbrowESR}$. Along the crystal symmetry axis z, the hyperfine structure anisotropy $(Fig.~\ref{CuAniso1},\ref{CuScratchPlot1}$ and $\ref{CuScratchPlot2})$ plays the roles with anisotropy energy$\citep{AbragamESR,PilbrowESR}$:
   
  \begin{eqnarray}
\label{eq:CuSH2}
\lefteqn ~ W = { P_{\scriptscriptstyle\parallel}\bigg\{ I_z^2 - \frac{1}{3} I(I+1) \bigg\} - \beta B {g^I_{\scriptscriptstyle\parallel Cu}}I_z}~~~~  
\end{eqnarray} 
 
  The condition $\beta B {g_{\scriptscriptstyle\parallel Cu}} \gg A_{\scriptscriptstyle\parallel Cu}\gg P_{\scriptscriptstyle\parallel} \gg\beta B {g^I_{\scriptscriptstyle\parallel Cu}}$ is implied for the hyperfine multiplet observations from a single spectroscopic term. Nuclear Zeeman energy term $\beta B {g^I_{\scriptscriptstyle\parallel Cu}}$ is naturally very small, and Bleaney et al$\citep{BleaneyIngram,BleaneyTrenam}$ measured $P_{\scriptscriptstyle\parallel}$ of $Cu^{2+}$ ion is about $0.0011~cm^{-1}$ in both tutton salts and lanthanum magnesium nitrate at $20~K$. We keep temperature $T\leq20~mK$ steadily maintaining a possible state of $k_B T < A_{\scriptscriptstyle\parallel Cu}$ for observation and measurement of hyperfine structure including anisotropies as nuclear dipolar hyperfine interaction.

\subsection{Aggregate frequency width of WG mode ESR spectrum and impurity ion concentration:}

\begin{figure}[h!]
\includegraphics[width=3.5in]{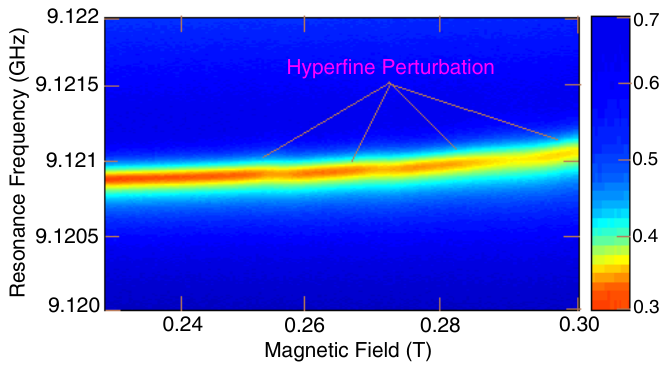}
\caption{\label{CuDens} Density plot of $Cu^{2+}$ ion ESR spectrum showing aggregate frequency width (width of deep red background) about $50kHz$ of aggregate spin ensemble around resonance frequency $9.121GHz$ of $WGH_{4,1,2}$ mode. Background frequency shift with the increase of magnetic field is due to local strain.}
\end{figure}

A light-yellow cylindrical $SrLaAlO_4$ crystal of height $9.04~mm$ and diameter $17.18~mm$ was inserted at the center of an oxygen-free cylindrical copper cavity. The crystal loaded cavity was cooled in a dilution refrigerator (DR) down to $20~mK$, with the low temperature improving the sensitivity as shown in Eq.\ref{eq:Nmin}. Fifteen WG modes with high-azimuthal-mode-number within a frequency range of $7~GHz$ to $18~GHz$, and thus electromagnetic energy filling factors of the order of unity were monitored. The applied DC-magnetic field is auto-controlled through the electric current to the superconducting magnet by computer program. The magnetic field within the range $-0.2~T$ to $1~T$ was applied ensuring by the computer in a step of sweep $4\times 10 ^{-4}~T$ maintaining stabilized temperature at $20~mK$. Each mode with frequency span $2~MHz$ was scanned in a very narrow frequency line width of a high resolution VNA for a period of five seconds at each step of magnetic field. The slow rate of sweep of magnetic field was applied keeping stablized temperature. The microwave input power of the SLA resonator was about $-60~dBm$. Mode volume of the order $10^{-7} m^3$, and $Q_L$ about 100,000 at $20~mK$. Aggregate frequency width $\Delta \omega \le 50~kHz$ is in the region $\omega\approx \omega_L$ for hybrid mode $WGH_{4,1,2}$ of resonance frequency $9.121~GHz$ and $Q_L$ about 75,000 (Fig.$\ref{CuDens}$). Also, the $WGH_{4,1,1}$ had$\citep{SLA}$ $\Delta \omega \le 20~kHz$ in the region $\omega\approx \omega_L$  of resonance frequency $9.072~GHz$ and $Q_L$ about 115,000. With a little dielectric variation among selected modes, observed $Q_L$ was always more than 50,000 at $20~mK$. In such a observed viable state, practically $\Delta \omega$ was less than line-width of all the selected WG modes. Hence the minimum number of ion (effective spin $S=\frac{1}{2}$) is required about $3\times 10^{10}$ within the WG mode volume at a noise level of ratio $\frac{P_n}{P}=1$ (see~Eq.$\ref{eq:Nmin}$). These adjustments ensure the sensitivity of this method at $20~mK$ in $SrLaAlO_4$ crystal of $Cu^{2+}$ ion allowing detection down to the $0.03~ppb$ level of concentration. 

For this technique, the factor $\frac{\Delta \omega}{\omega}$ in Eq.\ref{eq:Nmin} determines the required measurement accuracy\cite{CharlesESR} under influence of hyperfine coupling and effective spin S. Because, we can observe $\Delta\omega$ from the transmission spectrum density plot (Fig.\ref{CuDens}) and adjust the required Q-factor according to $\Delta\omega$. The number $N_{min}$ changes as a product with the term $\frac{3}{S(S+1)}$. On the other hand, in case of the calculation of $N_{min}$ as a notion of $\frac{spins}{G}$, the factor $\frac{\Delta B}{B}$ is used\cite{CharlesESR} instead of $\frac{\Delta \omega}{\omega}$ in the Eq.\ref{eq:Nmin}. Here $\Delta B$ is the line width, $B$ is the resonant magnetic field and $1G=10^{-4}T$.\\

To avoid the addition of thermal noise from room temperature, a $10~dB$ microwave attenuator was used at $4~K$ stage and another one at $1~K$ stage of the DR. Also, a $20~dB$ attenuator was added at $20~mK$ stage of the DR. These cold stage attenuation plus the use of a low noise temperature cryogenic amplifier at the output after the resonator ensures good enough signal to noise ratio $(SNR)$. From these ESR characteristics, we were able to identify the types of paramagnetic impurities present in the crystal with hyperfine couplings.

\subsection{Results and Discussion:}

\subsubsection{Hyperfine structure anisotropy:}

\begin{figure}[t!]
\includegraphics[width=3.3in]{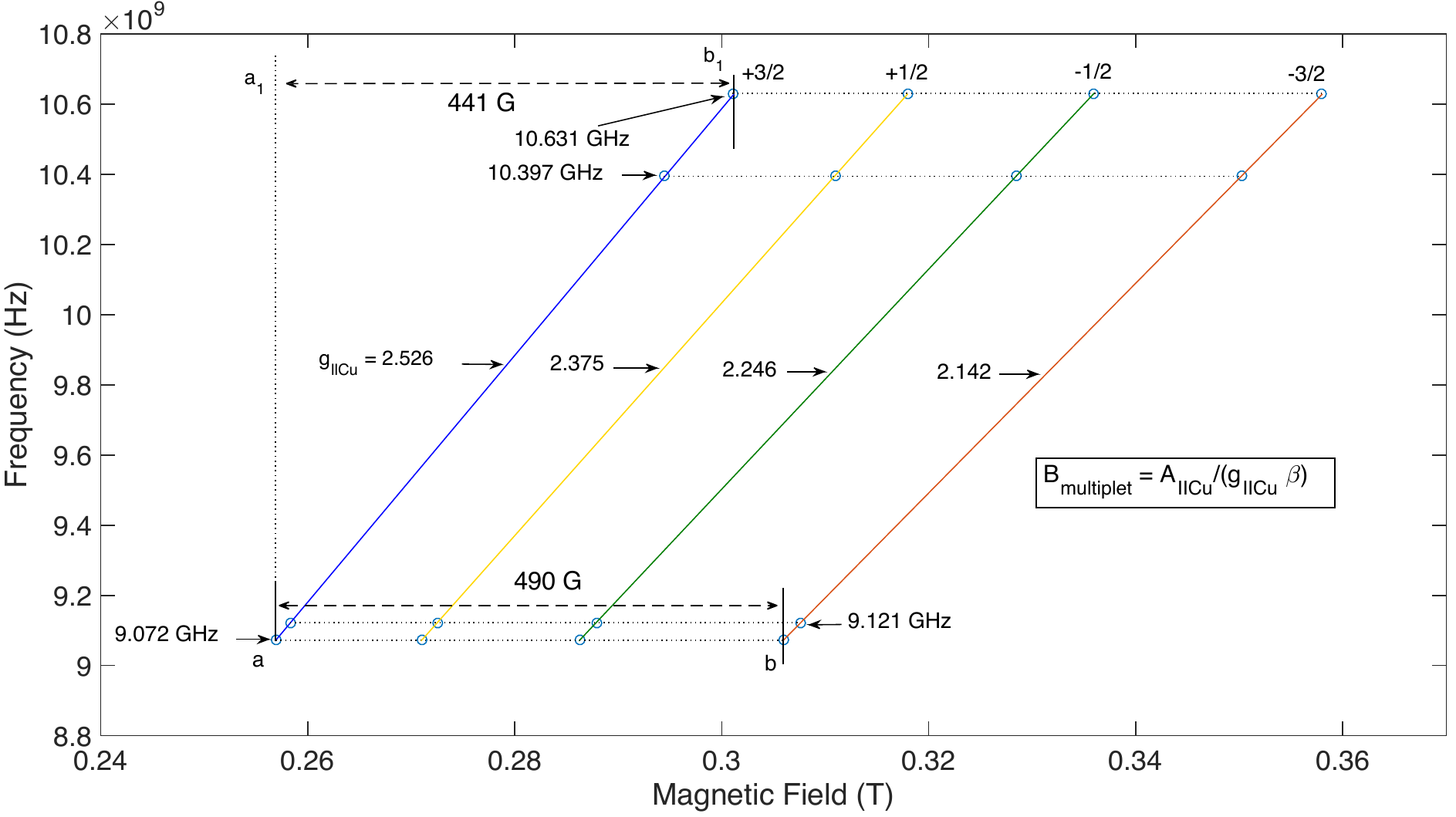}
\caption{\label{CuAniso1} g-factor and hyperfine structure anisotropy map of $Cu^{2+}$ ion with $WG$ modes of different frequency of ESR spectrum. The magnetic field under hyperfine coupling is $\textbf{B} = \textbf{B}_\circ - \frac{\textbf{A} M_I}{g_e\beta}$; and resonance frequency $\omega=-\gamma\textbf{B}_\circ$. The magnetic field $\textbf{B}_\circ$ is the external magnetic field of resonance without hyperfine coupling, and $\gamma$ is the gyromagnetic ratio of electron.}
\end{figure}

\begin{figure}[t!]
\includegraphics[width=3.3in]{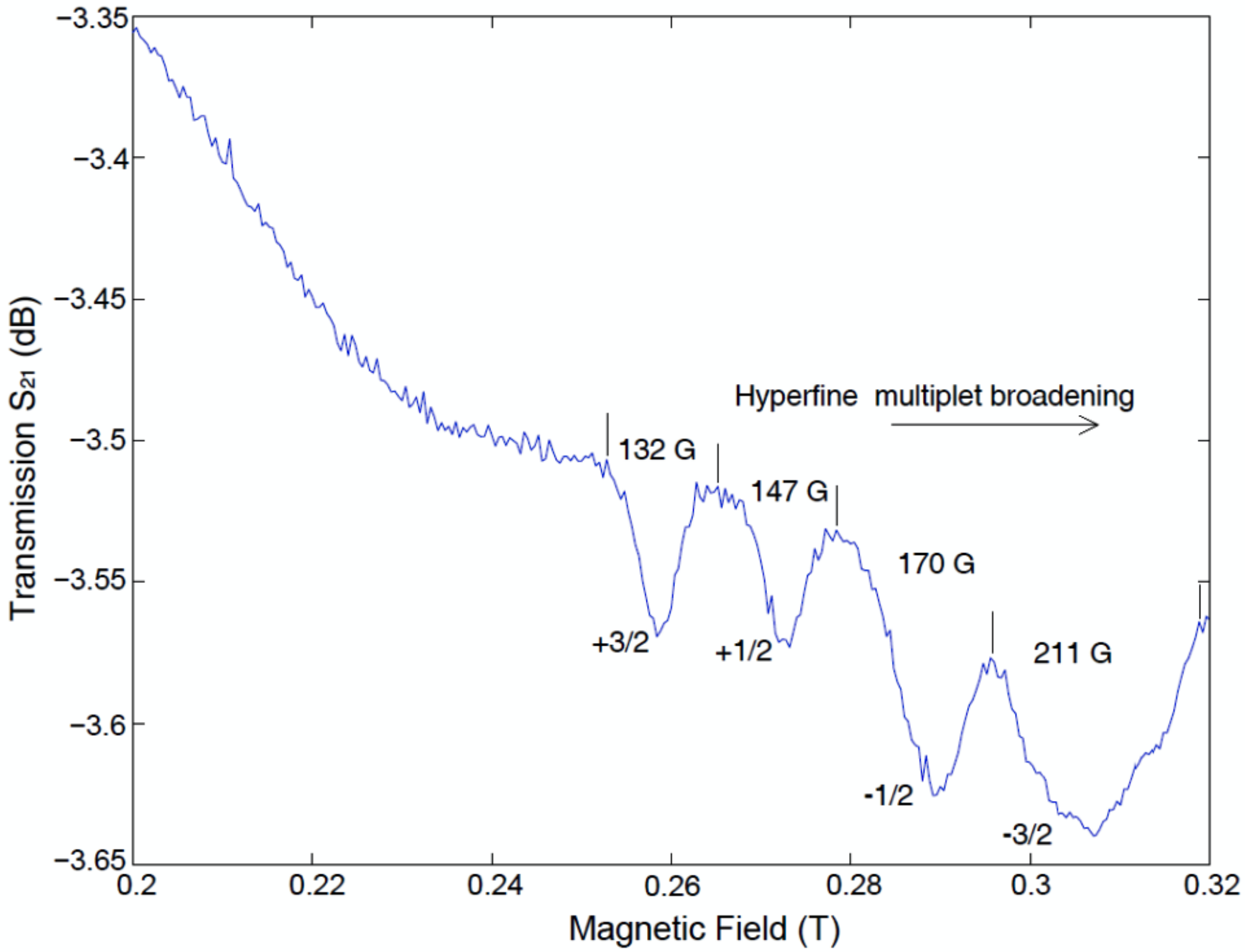}
\caption{\label{CuScratchPlot1} Hyperfine multiplet widths of $Cu^{2+}$ ion nuclear perturbation coupling with $WGH_{4,1,1}$ mode of frequency $9.072~GHz$ at $20~mK$.}
\end{figure}

The measured hyperfine multiplet of different resonance frequencies in a structure of four lines in the ESR spectrum map ($Fig.~\ref{CuAniso1}$) indicates an ion of $I=\frac{3}{2}$ which is the nuclear spin of $Cu^{2+}$ ion. It reveals hyperfine multiplets with a substantial anisotropy. The hyperfine structure width varies by $79~G$ from $132~G$ to $211~G$ as nuclear magnetic quantum number, $M_I$, varies from $+\frac{3}{2}$~to $-\frac{3}{2}$; and the broadening increases with the increase of applied magnetic field ($Fig.~\ref{CuAniso1}$ and $\ref{CuScratchPlot1}$). As a consequence, the measured parallel g-factors $g_{\scriptscriptstyle\parallel Cu}$ are of $2.526,~2.375,~2.246$ and $2.142$ ($Fig.~\ref{CuAniso1}$). The amount of hyperfine multiplet magnetic field in the $Fig.\ref{CuAniso1}$ are given as the hyperfine structure anisotropy of the four identified WG modes of resonance frequencies $9.072~GHz,~9.121~GHz,~10.397~GHz$ and $10.631~GHz$. For each $WG$ mode, it is showing $A_{\scriptscriptstyle\parallel Cu}$ variation between nuclear spin quantum numbers $+\frac{3}{2}$ to $-\frac{3}{2}$ $(Fig.~\ref{CuAniso1})$. These broadening effects are happening with the reduction of parallel g-factor between the steps of nuclear magnetic quantum numbers. In this perturbation, magnetic field varies according to $M_I$, and multiplet width may be written as$\citep{PilbrowESR}$~$\frac{A_{\scriptscriptstyle\parallel Cu}}{g_{\scriptscriptstyle\parallel Cu}\beta}$~$(Fig.\ref{CuAniso1}$ and $\ref{CuScratchPlot1})$. The observed variation of multiplet's magnetic field width among $,M_I,$ of $+\frac{3}{2}$ to $-\frac{3}{2}$ is apparently due to nuclear electric quadruple moment producing impact as dipolar hyperfine structure parameter $(P_{\scriptscriptstyle\parallel})$. These observations of spectroscopic terms are in a good agreement with theoretical models.\\

Basically, the formulas of spin-Hamiltonian parameters are considered with a typical value of $\frac{\lambda}{\Delta}\sim -0.05$ in the crystal field$\citep{AbragamESR}$. Here, $\lambda$ is the spin-orbit coupling constant and $\Delta=10D_q$ is the energy difference between $t_{2g}$ and $e_g$. But the correction term $\sim (\frac{\lambda}{\Delta})^2$ is not negligible for the most general case of a rhombic distortions$\citep{AbragamESR,BleaneyIngram,JT-Modern}$, which admixes the $e_g$ splitted doublet (splitting due to the tetragonal distortion) ground state $\mid x^2-y^2\rangle = \mid2^s\rangle$ with $\mid 3z^2-r^2\rangle = \mid 0\rangle$. For the ground state $\mid2^s\rangle$ in octahedral elongation along z-axis due to the tetragonal distortion, the parallel and perpendicular g-factors anisotropy are $\delta(g_{\scriptscriptstyle\parallel Cu})=8\times\frac{\lambda}{\Delta}$ and $\delta(g_{\scriptscriptstyle\perp Cu})=2\times\frac{\lambda}{\Delta}$ respectively (considering crystal field splitting between $e_g$ and $t_{2g}$ in average equal to $\Delta$)$\citep{BleaneyIngram,CuJT-Misra}$.\\

\begin{table}[t!]
\centering
\caption{\label{table1} Spin-Hamiltonian parameters of $Cu^{2+}$ ion in $SrLaALO_4$, and mean values for the two stable isotopes $^{63}Cu$, $^{65}Cu$ in $[La_2Mg_3(NO_3)_{12},(24H_2O~or~24D_2O)]$: Hyperfine structure constants $\textbf {A}$ and $\textbf {P}$ are in units of $10^{-4}~cm^{-1}$.}
\resizebox{\columnwidth}{!}{%
\begin{tabular}{lccccr}
\hline\hline
\textbf{Source}&\textbf{Temperature} &\textbf{$g$-factors}~~~&\textbf{A} and \textbf{P}\\ 
\hline {Theoretical\citep{DefectCu^2}} &  &$g_{\scriptscriptstyle\parallel Cu}=2.321$&$A_{\scriptscriptstyle\parallel Cu}= -150$\\
{($SrLaAlO_4$)}&  &$g_{\scriptscriptstyle\perp Cu}=2.070$&$A_{\scriptscriptstyle\perp Cu}=-3$\\
\hline {Experimental\citep{CuHole}}& $293~K$ & $g_{\scriptscriptstyle\parallel Cu}=2.321$&$A_{\scriptscriptstyle\parallel Cu}=-150$\\{($SrLaAl_{1-x}Cu_xO_4)$}& &$g_{\scriptscriptstyle\perp Cu}=2.069$&$A_{\scriptscriptstyle\perp Cu}=<10$\\
~~~~$x=0.02$\\
\hline {This Experiment}& 20~mK&$g_{\scriptscriptstyle\parallel Cu}=2.322$&$A_{\scriptscriptstyle\parallel Cu}=-174.6$\\
 $(SrLaAlO_4)$& &$g_{\scriptscriptstyle\perp Cu}=2.053$&$A_{\scriptscriptstyle\perp Cu}=13.4$~~\\
& &$\frac{1}{3}(g_{\scriptscriptstyle\parallel Cu} + 2 g_{\scriptscriptstyle\perp Cu})$& $\frac{1}{3}(A_{\scriptscriptstyle\parallel Cu} + 2 A_{\scriptscriptstyle\perp Cu})$\\
& &$=2.142(6)$ &$=-49.26$\\
& & & $P_{\scriptscriptstyle\parallel}=+12.3$\\
\hline{Experimental$\citep{BleaneyIngram}{^,}\citep{BleaneyTrenam}$}& $90~K$&$g_{\scriptscriptstyle\parallel Cu}=2.219(3)$&$|A_{\scriptscriptstyle\parallel Cu}|=29.0(5)$\\
$[La_2Mg_3(NO_3)_{12}]$& & $g_{\scriptscriptstyle\perp Cu}=2.218(3)$& $|A_{\scriptscriptstyle\perp Cu}|=27.5(5)$\\
\hline{Experimental$\citep{JT-HyperBroad}{^,}\citep{BleaneyTrenam}$}&$20~K$& $g_{\scriptscriptstyle\parallel Cu}=2.465(1)$&$A_{\scriptscriptstyle\parallel Cu}=-111.7(5)$\\
$[La_2Mg_3(NO_3)_{12}]$& &$g_{\scriptscriptstyle\perp Cu}=2.099(1)$&$|A_{\scriptscriptstyle\perp Cu}|=16.0(5)$\\
& &$\frac{1}{3}(g_{\scriptscriptstyle\parallel Cu} + 2 g_{\scriptscriptstyle\perp Cu})$& $|\frac{1}{3}(A_{\scriptscriptstyle\parallel Cu} + 2 A_{\scriptscriptstyle\perp Cu})|$\\
& &$=2.221(1)$ & $=26.6(5)$\\
& & & $P_{\scriptscriptstyle\parallel}=+10.5(5)$\\
\hline
\end{tabular}
}
\end{table}

\begin{figure}[t!]
\includegraphics[width=3.1in]{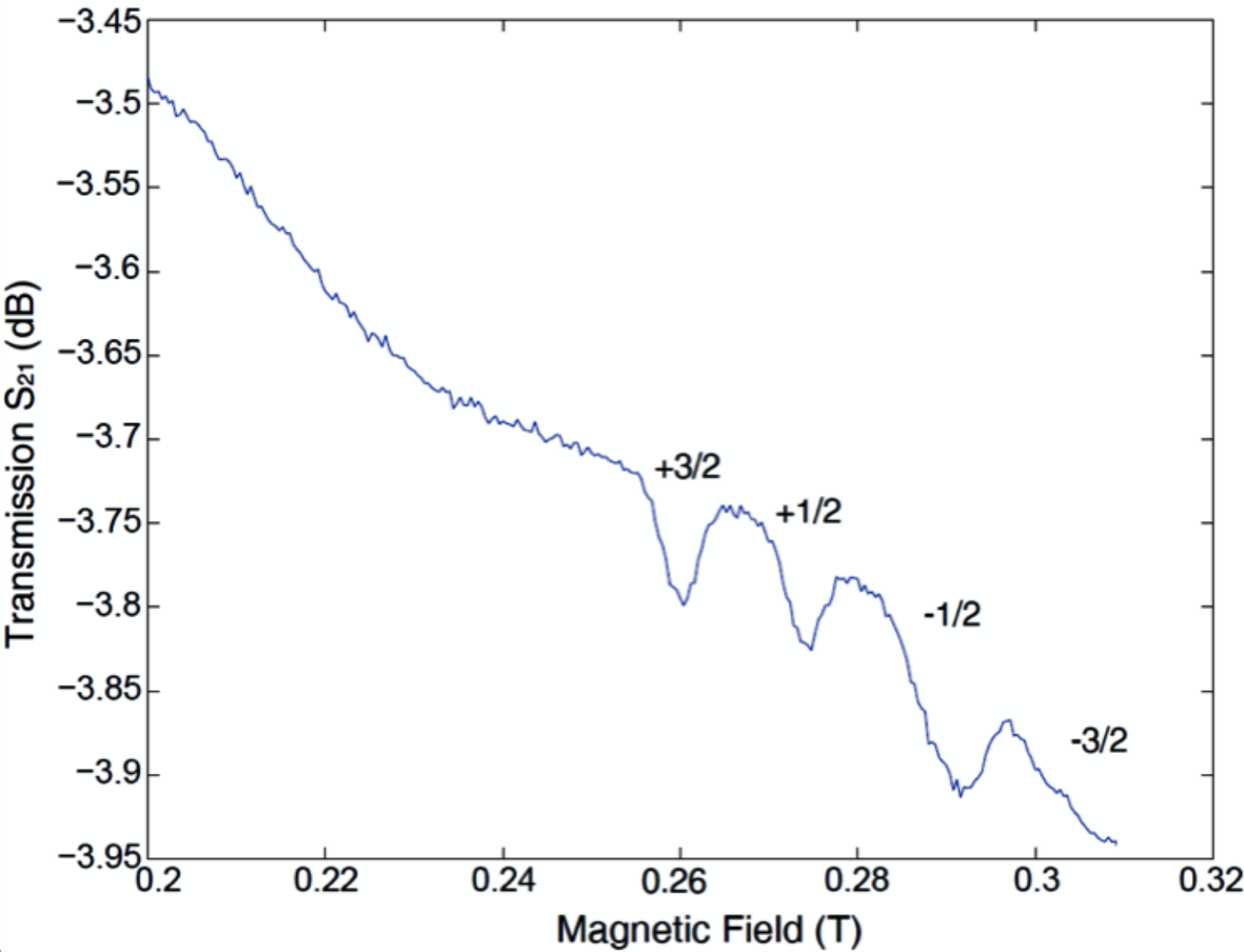}
\caption{\label{CuScratchPlot2} Hyperfine multiplet widths of $Cu^{2+}$ ion nuclear perturbation coupling with $WGH_{4,1,2}$ mode of frequency $9.121~GHz$ at $20~mK$). Transmission spectrum disappeared at the position of $M_I=-\frac{3}{2}$ (magnetic field $0.31~T$) due to local strain in the $CuO_6$ octahedral structure.}
\end{figure} 

Taking average of the measured parallel g-factors, the parallel g-factor anisotropy $\delta(g_{\scriptscriptstyle\parallel Cu})$ is about $0.32$. Using the values of measured $g_{\scriptscriptstyle\parallel Cu}$, calculated perpendicular g-factor anisotropies are $\delta(g_{\scriptscriptstyle\perp Cu})\simeq 0.051$ in $SrLaAlO_4$ crystal. From the formula, taking $\Delta \sim 12,300~cm^{-1}$, these g-factor anisotropy values $\sim0.4$ of parallel g-factor and $\sim0.1$ of perpendicular g-factor were measured by Bleaney et al$\citep{BleaneyIngram,BleaneyTrenam}$. They considered $Cu^{2+}$ ion in a crystal like lanthanum magnesium nitrate salt where $Cu^{2+}$ is at the site of substituted diamagnetic $Mg^{2+}$ ion (see~Table-\ref{table1}). The measured values of g-factor anisotropy in $SrLaAlO_4$ are slightly smaller than their measured values of g-factor anisotropy, plausibly the cause of co-valency effect in the divalent $Cu^{2+}$ ion site of the substituted trivalent $Al^{3+}$ ion. It may be mentioned that $Al^{3+}$ ion is not diamagnetic, and its outermost electron shell is $p$-orbital which is quenched in ESR transitions. Also, the measured values agreed with the theoretical condition $g_{\scriptscriptstyle\parallel Cu}>g_{\scriptscriptstyle\perp Cu}>g_s$, which is implied for the elongation along $z$-axis ($g_s$ is the free electron g-factor).\\

 In the $O^{2-}$ ions coplanar bonds in four fold axis (of four $Cu-O1$ bonds and two elongated $Cu-O2$ bonds), a hole de-localization due to metal-ligand charge transfer is considered $\citep{CuHole}$. Divalent copper oxidation stage $+2.46$ in the substituted trivalent $Al^{3+}$ ion site supports this arguments as a consequence of partial hole contribution$\citep{CuHole}$. Iodometric titration gives  this average value of the copper oxidation (of $+2.46$) for the formatted hole in $CuO_6$ center of $Al_{1-x}Cu_xO_2$ with $x=0.2, 0.4, 0.6$ and $0.8$ solid solution$\citep{CuHole1}$. Yu. V. Yablokov et al.$\citep{CuHole}$ observed that the lattice constant, c, is strongly influenced by the copper concentration. They found that in $LaSrAl_{1-x}Cu_xO_4$, for $x=1$ the value $c=12.97$~\r{A} virtually coincids with the value $13.11$~\r{A}.\\
 
  Hence, naturally the substitution of divalent $Cu^{2+}$ ion in trivalent $Al^{3+}$ ion site changes the stoichiometry and has significant effect on ESR spectrum, but the influence of concentration has been disregarded for such a low concentration level (of the order of ppb) of impurity $Cu^{2+}$ ion in case of the crystal $SrLaAlO_4$. Since the ionic radius of copper ($r_{\scriptscriptstyle Cu}=0.72$~\r{A}) is larger than the substituted aluminium ion radius ($r_{\scriptscriptstyle Al}=0.51$~\r{A}), Wei et al.$\citep{DefectCu^2}$ found that this causes a displacement making the $Cu-O1$ bond length $R_\perp'=1.945$~\r{A} shifting away the oxygen $0.06$~\r{A} from the $Cu^{2+}$ ion. The increase of any copper-oxygen bond length of $CuO_6$ octahedral has a vital role in the formation of Jahn-Teller effect.\\
 
\begin{figure}[t!]
\includegraphics[width=3.3in]{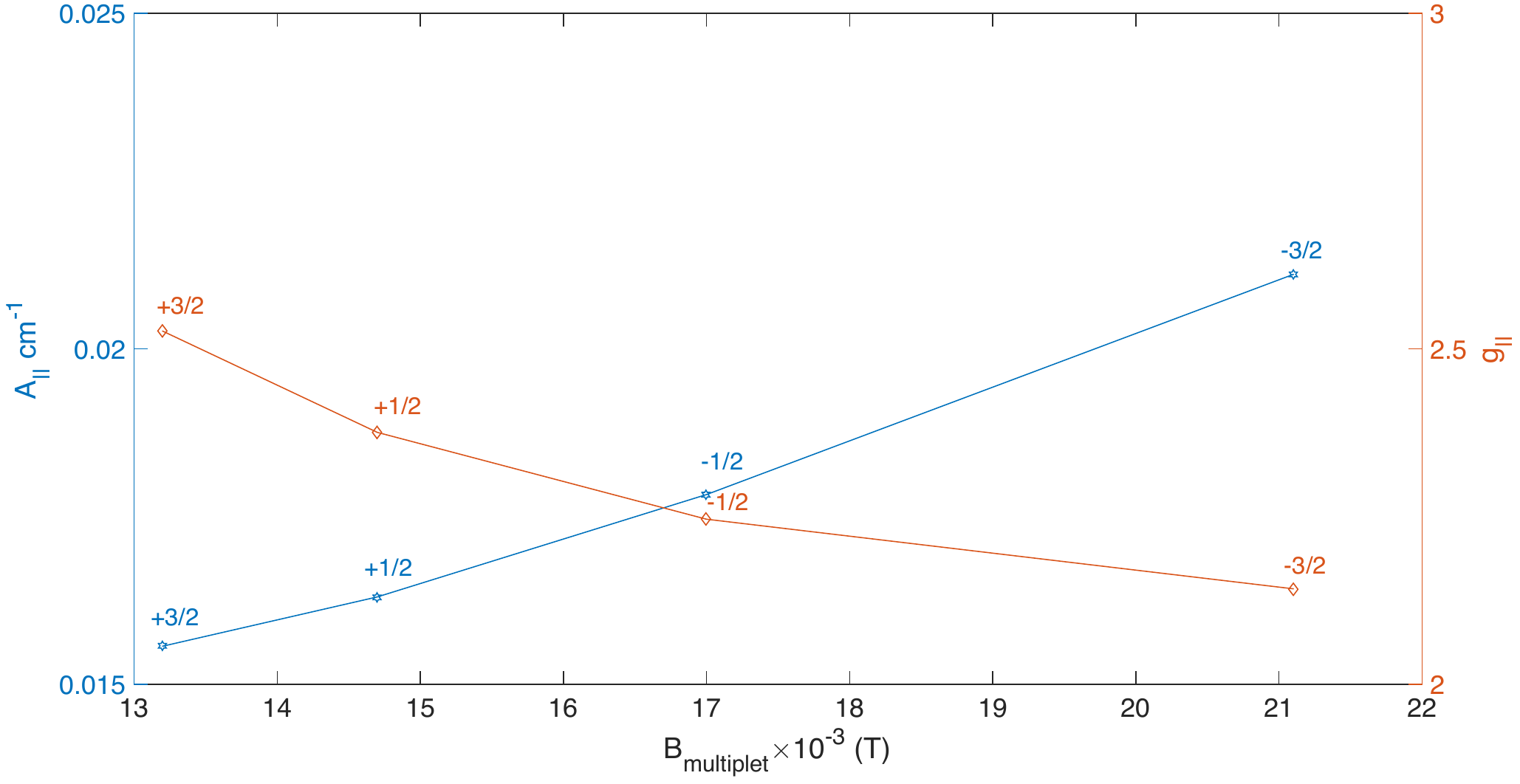}
\caption{\label{CuAnisoAgB} g-factor anisotropy and hyperfine structure anisotropy of $Cu^{2+}$ ion in nuclear perturbation.}
\end{figure} 

 In the expressions for the hyperfine structure, the terms involving $\frac{\lambda}{\Delta}$ has been evaluated from the measurements of the g-factor anisotropies in the multi mode ESR spectrum mapping $(Fig.\ref{CuAniso1})$. Measured values of $A_{\scriptscriptstyle\parallel Cu}$ and calculated $A_{\scriptscriptstyle\perp Cu}$ can be regarded as determining two unknowns, $\langle r^{-3} \rangle$ and the core polarization parameter $,k,$ using available values of $g^I_{\scriptscriptstyle\parallel Cu}$. To explain the hyperfine structure broadening, the second order correction term $\sim (\frac{\lambda}{\Delta})^2$ is useful for the case of a rhombic distortion which admixes the ground state $\mid x^2-y^2\rangle = \mid2^s\rangle$ with $\mid 3z^2-r^2\rangle = \mid 0\rangle$ and broaden the hyperfine structure about $79~G$.\\

 An angle, $\phi$, is corresponded to a lower symmetry, and the corresponding eigenstates are; ground state $cos\frac{\phi}{2}\mid x^2-y^2\rangle + sin\frac{\phi}{2}\mid 3z^2-r^2\rangle$ with the orthogonal excited state $sin\frac{\phi}{2}\mid x^2-y^2\rangle - cos\frac{\phi}{2}\mid 3z^2-r^2\rangle$ which is higher in energy by $4\times (Jahn-Teller~energy~ W_{JT})\citep{AbragamESR,JT-Modern,JT-Static1}$. Hence, the admixture of the excited state is given by $sin^2(\frac{\phi}{2})$ for this lower symmetry distortion$\citep{CuJT-Misra}$. From this experiment results, it may be estimated as $\phi\sim tan^{-1}(\frac{79}{132+147+170+211})=6.82^\circ~(Fig.\ref{CuScratchPlot1})$, and which means that the admixture is about $0.36\%$. This estimation of angle is nearly equal to the calculated angle using the metal-ligand bond lengths of $CuO_6$ octahedral structure.

\subsubsection{Jahn-Teller effect and parameters:}

The observed broadening $79~G$ of the $Cu^{2+}$ ion hyperfine structure or multiplet($Fig.~\ref{CuAniso1},\ref{CuScratchPlot1}$ and $\ref{CuScratchPlot2}$) along the increase of magnetic field may be explained by nuclear quadruple interaction in the symmetry of the ligand field of $CuO_6$ octahedral in the $SrLaAlO_4$ crystal. Nuclear displacement of copper lowers the symmetry of $CuO_6$ producing a special coupling between the electronic and nuclear motion$\citep{JT-Modern}$.\ 

In the distorted rhombic lower symmetry of $CuO_6$ structure at $20mK$ supports the occurrence of a static Jahn-Teller effect $(SJTE)$. Burns and Hawthorne$\citep{JT-Static}$ showed that the state of lowest energy in which octahedron suffers an elongation in tetragonal symmetry are corresponded to a $SJTE$. Hence with the elongation in the octahedron ligand complex $CuO_6$ under rhombic deformation in $SrLaAlO_4$ tetragonal crystal, the variation of measured hyperfine multiplet widths $-155.7\times10^{-4}~cm^{-1}~(13.2~mT),~ -163.0\times10^{-4}~cm^{-1}~(14.7~mT),~ -178.3\times10^{-4}~cm^{-1}~(17.0~mT)$ and $-211.1\times10^{-4}~cm^{-1}~(21.1~mT)$ with the order of nuclear magnetic quantum number $+\frac{3}{2},~+\frac{1}{2},~-\frac{1}{2},$ and $-\frac{3}{2}$ respectively at $9.072~GHz~(WGH_{4,1,1})$ $(Fig.~\ref{CuScratchPlot1}$) can be referred to the effect of $SJTE$.\\
 
Breen, Krupka and Williams$\citep{JT-HyperBroad}$ found that the spin-lattice relaxation time $(\tau)$ at liquid helium temperature is about four orders of magnitude faster for $Cu^{2+}$ in lanthanum magnesium nitrate than in tutton salt of tetragonal or lower symmetry of ligand field. They relate this result to the rate of relaxation $(\frac{1}{\tau})$ between equivalent distortions without reorientation of electron spin. Similarly, in continuous cooling, maximum Q-factor of WG modes were attained around $10K$ and decreased about ten times in magnitude at $20mK$ in $SrLaAlO_4$. It proves the increase of relaxation time at $20~mK$.\

\begin{figure}[b!]
\includegraphics[width=3.3in]{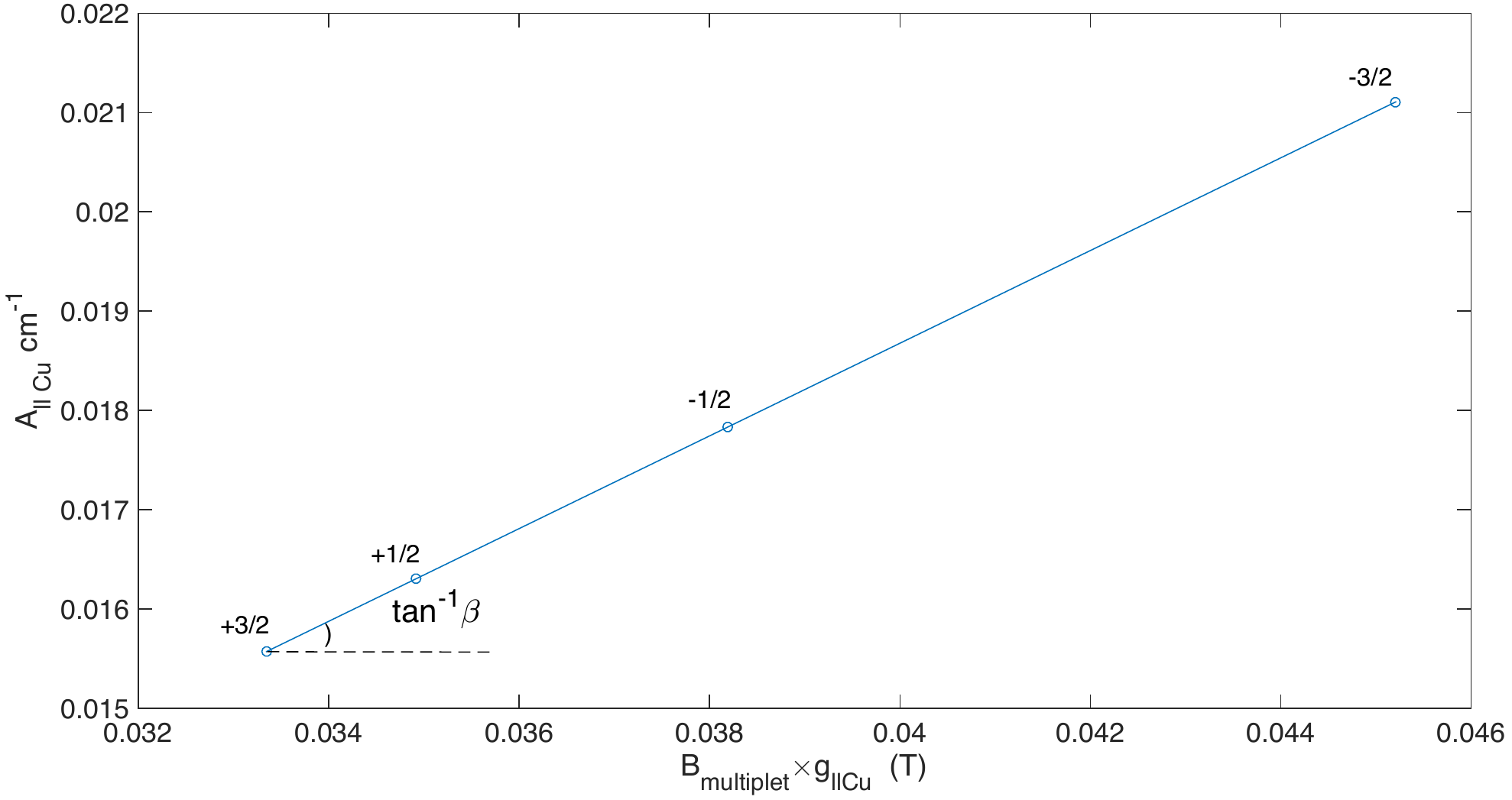}
\caption{\label{CuAnisoBohr} Bohr magneton is revealed in hyperfine structure anisotropy of $Cu^{2+}$ ion.}
\end{figure}

\begin{figure}[b!]
\includegraphics[width=3.6in]{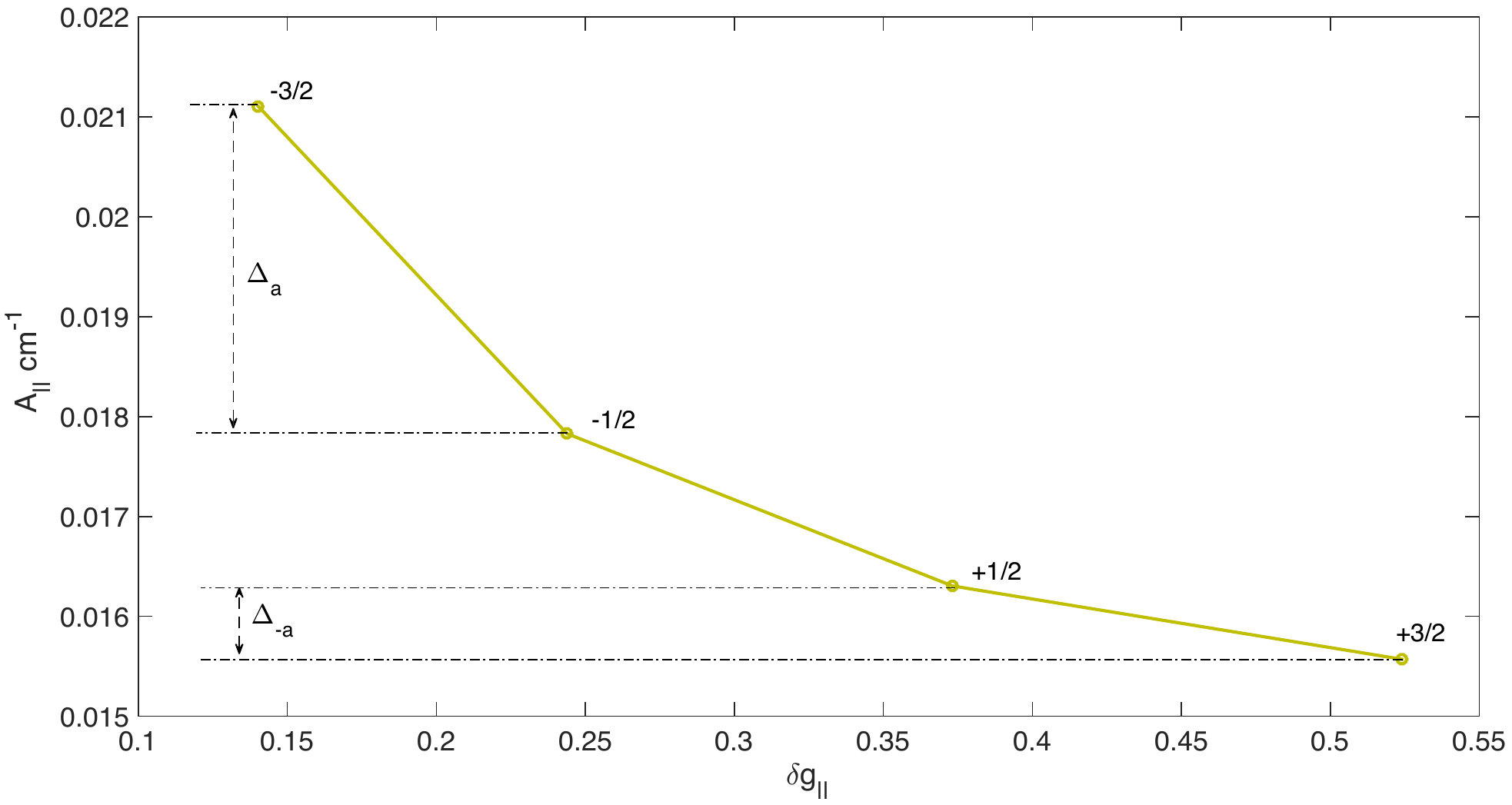}
\caption{\label{CuAnisoA} Impacts of nuclear dipolar hyperfine structure parameter of $Cu^{2+}$ ion in the substituted $Al^{3+}$ ion site.}
\end{figure}

Generally, a rhombic distortion is corresponded to principle g-factors in the direction of three mutually perpendicular four-fold axes of $CuO_6$. These g-factors to the order $\frac{\lambda}{\Delta}$ are$\citep{AbragamESR}$:
\begin{eqnarray}
\label{eq:CuJT}
 {{g_1}= g_s - \frac{2\lambda}{\Delta}\{cos\frac{1}{2}\phi - \sqrt{(3)}sin\frac{1}{2}\phi\}^2 \nonumber}\\ 
 {{g_2}= g_s - \frac{2\lambda}{\Delta}\{cos\frac{1}{2}\phi + \sqrt{(3)}sin\frac{1}{2}\phi\}^2 \nonumber}\\
{g_3}= g_s - \frac{8\lambda}{\Delta}\{cos^2\frac{1}{2}\phi\}^2    ~~~~~~~~~~~~~~~~~
\end{eqnarray}

 Here, $\phi$ is the vectorial angle of a polar coordinate system $(\rho,\phi)$ which describes the $\Gamma_3(E)$ distortions corresponding to the normal-coordinates transformation $Q_\varepsilon (=\rho~sin~\phi)$ and $Q_\theta (=\rho~cos~\phi)$. At the experiment temperature $20~mK$, g-factors along four-fold axes in the octahedral symmetry is corresponded to $g_3>g_2>g_1$ with the elongation along z-axis. In such a lower symmetry in the Jahn-Teller stabilization energy of the $CuO_6$ metal ligand system, an admixture of the ground state $\mid x^2-y^2\rangle$ with the excited $\mid 3z^2-r^2\rangle$ state is evident under the condition $R=\frac{g_2-g_1}{g_3-g_2}<1$ as implied for the elongation along z-axis$\citep{CuJT-Misra}$. The observed hyperfine structure anisotropy follows the equation no.$\ref{eq:CuSH2}$ under condition $A_{\scriptscriptstyle\parallel Cu}\gg P_{\scriptscriptstyle\parallel}$. In the regular ESR spectrum of $Fig.~\ref{CuScratchPlot2}$, the multiplet at $M_I=-\frac{3}{2}$ disappeared clearly due to local strain effect. Wei et al.$\citep{DefectCu^2}$ found that the $Cu-O1$ bond length increases shifting away the oxygen $0.06$~\r{A} from the $Cu^{2+}$ ion. From this significant amount of covalent bond length extension, it is plausible to say that the perturbation due to local strain is large in comparison with Zeeman anisotropy energy difference $(g_{\scriptscriptstyle\parallel Cu}-g_{\scriptscriptstyle\perp Cu})\beta B$.\\
 
 The variation characteristics of $g_{\scriptscriptstyle\parallel Cu}$ and $A_{\scriptscriptstyle\parallel Cu}$ with respect to multiplet width in magnetic field, $B_{multiplet}$, has been shown in the $Fig.~\ref{CuAnisoAgB}$ with corresponding $M_I$. With the measured parameters $A_{\scriptscriptstyle\parallel Cu}$, $g_{\scriptscriptstyle\parallel Cu}$ and multiplet width in terms of magnetic field according to $M_I$, we found Bohr magneton $\beta=9.23\times 10^{-24}JT^{-1}$ from the graph in the Fig.~$\ref{CuAnisoBohr}$. The characteristic straight line supports the theoretical predictions and consistency of this measurement. According to the equation no.$\ref{eq:CuSH2}$, the parameter, $P_{\scriptscriptstyle\parallel}=\frac{\bigtriangleup_{a} - \bigtriangleup_{-a}}{2}=12.3\times10^{-4}~cm^{-1}$, has been measured as an anisotropy of $A_{\scriptscriptstyle\parallel Cu}$ due to nuclear quadruple moment $(Fig.~\ref{CuAnisoA})$. This value of $P_{\scriptscriptstyle\parallel}$ reveals $\langle r_q^{-3}\rangle = 5.23~a.u.$ with nuclear quadruple moment $Q=-0.211~barn\citep{NuclearQ}$. $\langle r^{-3}\rangle\simeq7.5$ a.u. for free cupric ion, and  6.3 a.u. is measured value of this ion in lanthanum magnesium nitrate$\citep{BleaneyPryce}$. These values supports that $\langle r_q^{-3}\rangle = 5.23~a.u.$ in $SrLaAlO_4$ is valid.\\

\subsection{Conclusion:}
High WG mode $Q$-factor in this crystal at $\leq20~mK$ has allowed sensitive multi-mode ESR, and measurement of hyperfine structure anisotropy of spin along with site symmetry. The temperature was kept at $\leq 20~mK$ to observe the hyperfine multiplets. Rapid broadening of $Cu^{2+}$ ion hyperfine structure due to static Jahn-Teller effect $\cite{CuJT-Misra,JT-Static,JT-HyperBroad}$ are underlined of nuclear quadruple moment in a lower symmetry of paramagnetic ion's site in the elongated $CuO_6$ octahedral structure$\citep{JT-Ferro,JT-Modern}$. Ground state becomes unstable due to displacement of the copper nucleus in the centro-symmetric octahedron$\citep{JT-Ferro,JT-Permittivity}$. At least two electronic states of the reference configuration should be involved in rationalization of structural changes in the polyatomic system. JTE is a transformation in the two electron state paradigm in formation, deformation, and transformation of molecular systems and solids$\citep{JT-Green,JT-Review,JT-flex}$. The hyperfine multiplet anisotropy measurement is valid only if $|\frac{P_{\scriptscriptstyle\parallel}}{A_{\scriptscriptstyle\parallel Cu}}|$ is quite small than unity. Determination of hyperfine line spacing and broadening as well as the justification of covalency contributes to the microscopic state analysis at millikelvin temperatures. This WG mode ESR spectroscopy shows an alternative process for the estimation of Bohr magneton and $\langle r_q^{-3}\rangle$. The results are noteworthy as the first example of nuclear quadruple interaction observation as an anisotropy of hyperfine structure using microwave WG multi mode ESR.
                                                      
\begin{acknowledgements}

This work was funded by Australian Research Council (ARC) Grant no.CE110001013. Thanks to  Dr. Warrick G. Farr for assistance with data acquisition and Mr. Steve Osborne for technical support. 
\end{acknowledgements}

\subsection{References:}

\end{document}